\documentclass[12pt,preprint]{aastex}

\shorttitle{The Counter-Streaming Instability in Dwarf Ellipticals with
  Off-Center Nuclei}

\shortauthors{De Rijcke \& Debattista}

\begin{document}

\title{The Counter-Streaming Instability in Dwarf Ellipticals with Off-Center
  Nuclei\altaffilmark{1}}

 \author{Sven De Rijcke\altaffilmark{2}} \affil{Sterrenkundig
    Observatorium, Universiteit Gent, Krijgslaan 281, S9, B-9000 Gent,
    Belgium} \email{sven.derijcke@UGent.be} \author{Victor~P. Debattista}
    \affil{Institut f\"ur Astronomie, ETH H\"onggerberg, HPF G4.2,
    CH-8093 Z\"urich, Switzerland} \email{debattis@phys.ethz.ch}
    \altaffiltext{1}{Based partly on observations collected at the European
    Southern Observatory, Paranal, Chile (ESO Large Program
    165.N~0115)} \altaffiltext{2}{Research Postdoctoral Fellow of the
    Fund for Scientific Research - Flanders (Belgium)(F.W.O.)}

\begin{abstract}
  In many nucleated dwarf elliptical galaxies (dE,Ns), the nucleus is
  offset by a significant fraction of the scale radius with respect to
  the center of the outer isophotes.  Using a high-resolution $N$-body
  simulation, we demonstrate that the nucleus can be driven off-center
  by the $m=1$ counter-streaming instability, which is strong in
  flattened stellar systems with zero rotation. The model develops a
  nuclear offset of the order of one third of the exponential
  scale-length.  We compare our numerical results with photometry and
  kinematics of FCC046, a Fornax cluster dE,N with a nucleus offset by
  1\farcs2; we find good agreement between the model and FCC046. We
  also discuss mechanisms that may cause counter-rotation in dE,Ns and
  conclude that the destruction of box orbits in an initially
  triaxial galaxy is the most promising.
\end{abstract}

\section{Introduction}
\label{sec:intro}

Dwarf elliptical galaxies (dEs), faint, low-surface brightness
galaxies with smooth elliptical isophotes, are the most common type of
galaxy in clusters and groups of galaxies (see \citet{bf94}).
In hierarchical models of cosmological structure formation, dwarf
galaxies form first and subsequently merge to form larger
galaxies. Understanding their formation and evolution is therefore
vital to a complete picture of structure formation. There is evidence
that dEs constitute a heterogenous class of objects, kinematically
(rotating versus unrotating, \citet{ggv03}), chemically
\citep{mddzh03} and morphologically \citep{bbj02}. About half of the
dEs harbor a central bright nucleus and are called nucleated dEs
(dE,Ns). dE,Ns appear to be significantly older than normal
ellipticals and non-nucleated dEs \citep{rs03}; upon tidal stripping,
their nuclei have been suggested as sources of both the recently
discovered ultra compact dwarfs \citep{p01} and massive globular
clusters like $\omega$Cen \citep{g02}. In $\sim 20\%$ of Virgo dE,Ns,
the nucleus is significantly displaced with respect to the center of
the outer isophotes, typically by $\sim 1''$ ($\sim 100$~pc)
\citep{bbj00}. There is a tendency for the displacement to increase
with decreasing surface brightness but no relation between nuclear
displacement and any other structural or environmental parameter has
been found.

Various models have been proposed to explain these offset nuclei. In
this Letter, we investigate whether the lopsided ($m=1$)
counter-streaming instability can reproduce a dE,N with an off-center
nucleus. This instability has been known since \citet{zh78}, and been
studied analytically \citep{s88,pp90} and with $N$-body simulations
\citep{ms90,lds90,sm94,sv97}. \citet{sv97} do not detect it in systems
rounder than E6, which is mainly due to their rounder models being
stabilized by a higher radial pressure. On the other hand,
\citet{ms90} find lopsidedness developing in systems as round as E1
but with negligible radial pressure. Partial rotation only introduces
a pattern speed in an otherwise purely growing instability;
simulations find that the lopsidedness produced by the instability is
long-lived.

We use a realistic multi-component $N$-body model to generate a lopsided
system which we then compare with observations. In Section \ref{sec:theo}, we
present the $N$-body model and in Section \ref{sec:obs} we compare it with
photometry and kinematics of FCC046, an example of such a dE,N with an offset
nucleus. In Section \ref{sec:disc}, we discuss ways in which counter-rotation
may arise in dE,Ns.

\section{$N$-body simulation} 
\label{sec:theo}

We have been using $N$-body simulations to explore the
counter-streaming instability as a mechanism for producing lopsided
dE,Ns. 
We have performed both fully self-consistent simulations with live
nucleus, disk and halo run on a tree code, and restricted simulations
with rigid halos run on a grid code; both types of simulations will be
presented elsewhere. Here we present only one simulation which happens
to match FCC046 best, consisting of live disk and nucleus components
inside a rigid halo. The rigid halo was represented by a spherical
logarithmic potential with core-radius $r_{\rm c}$ and asymptotic circular
velocity $v_0$.  The initial disk was modeled by an exponential disk
of mass $M_{\rm d}$, scale-length $R_{\rm d}$, Gaussian thickness
$z_{\rm d}$ and truncated at $R_{\rm t}$.
The nucleus of mass $M_{\rm n}$ was generated by iteratively
integrating a distribution function in the global potential
\citep{pt70}. We used a distribution function of the form
$f(\vec{x},\vec{v}) \propto \left\{ \left[-E(\vec{x},\vec{v})
  \right]^{1/2} - \left[-E_{\rm max}\right]^{1/2} \right\}$. Here
$E_{\rm max} = \Phi_{\rm tot}(r_{\rm n})$, the total potential at the
nucleus truncation radius, $r_{\rm n}$, in the disk mid-plane.
Initial disk kinematics were set up using the epicyclic approximation
to give Toomre $Q = 1.3$. Vertical equilibrium was obtained by
integrating the vertical Jeans equation. The disk and nucleus were
represented by $36.8 \times 10^5$ and $3.2 \times 10^5$ equal-mass
particles, respectively. After setup, we switched the direction of the
velocity for half the particles, producing an unrotating disk and a
rotating nucleus. We used a quiet start \citep{s83} to cancel spurious
momenta and angular momenta.

In our units, where $R_{\rm d} = M(=M_{\rm n}+M_{\rm d}) = G = 1$,
which give a unit of time $(R_{\rm d}^3/GM)^{1/2}$, we set $z_{\rm d}
= 0.2$, $R_{\rm t} = 5$, $r_{\rm n} = 0.78$, $r_{\rm c} = 5$ and $v_0
= 0.648$.  The simulation was run on a 3-D cylindrical polar grid code
(described in \citet{sv97}) with $N_R\times N_\phi \times N_z = 60
\times 64 \times 225$.  The vertical spacing of the grid planes was
$\delta z = 0.0125$. We used Fourier terms up to $m=8$ in the
potential, which was softened with the standard Plummer kernel, of
softening length $\epsilon = 0.0125$. Time integration was performed
with a leapfrog integrator using a fixed time-step $\delta t = 0.006$.
To scale units to FCC046, we note that its surface-brightness profile
declines exponentially with scale-length $R_{\rm d}=4\farcs1\approx
0.4$~kpc. We estimated the stellar mass of FCC046 at $M \approx 1.0
\times 10^9 M_\odot$, given a $B$-band total luminosity $L_B = 2
\times 10^8 L_{B,\odot}$ and using a $B$-band mass-to-light ratio
$M/L_B \approx 5 M_\odot/ L_{B,\odot}$, typical for an old (15~Gyr),
rather metal-poor ([Fe/H]$=-0.5$) stellar population.  With this
scale-length and mass, the unit of time corresponds to 
$(R_{\rm d}^3/GM)^{1/2} \approx 4$~Myr and the velocity unit becomes
$(GM/R_{\rm d})^{1/2} \approx 100$~km/s.

\clearpage

\begin{figure}
\vspace{8cm}
\special{hscale=53 vscale=53 hsize=570 vsize=240 
         hoffset=-35 voffset=-165 angle=0 psfile="f1.eps"}
\caption{Panel {\bf a}:~a $B$-band image of FCC046; panel {\bf b}:~the
surface brightness of the $N$-body model at $t=150$. The isophotes are
spaced by 1~mag/arcsec$^2$ with the $\mu_B=24$~mag/arcsec$^2$ isophote
drawn in white. Panels {\bf c} and {\bf d}:~the model's velocity field
and the velocity dispersion field, respectively. The velocity field
perturbations have an $m=2$ symmetry and range between $-0.06$ (black)
and $+0.06$ (white), corresponding to $\pm 0.1 \times v_{\rm circ}$.
\label{modelps}}
\end{figure}

\clearpage

The global lopsidedness of the model, as measured by the amplitude of
the $m=1$ term in a Fourier expansion of the surface density, rose
exponentially with time, saturating at $t\simeq 70$, for an
$e$-folding time $\tau = 5.67$. 
The angular momentum of the disk and the orbital angular momentum of
the nucleus increase at the expense of the nucleus's internal
(rotational) angular momentum. By the end of the simulation, the
nucleus has lost approximately 20~\% of its initial internal angular
momentum. The resulting orbital velocity of the nucleus ($v_{\rm nuc}
\approx 0.01$) is, however, only a few percent of the local circular
velocity ($v_{\rm circ} \approx 0.35$), so that dynamical friction
against a live dark matter halo is not likely to damp the instability,
as is also borne out by the live halo simulations.

\section{FCC046:~a case study} 
\label{sec:obs}

FCC046 is an $m_B = 15.99$ dE4,N in the Fornax cluster.  Photometry
and long-slit spectra, with an instrumental broadening $\sigma_{\rm
instr}=30$~km/s, in the wavelength region containing the strong
absorption lines of the near-infrared Ca{\sc ii} triplet, were
observed at the VLT with FORS2 in November 2001 (photometry) and 2002
(spectroscopy). Seeing conditions were typically 0\farcs8 FWHM. A
detailed analysis of the photometry has been presented in
\citet{ik03b}, while details of the kinematics of FCC046 will be
presented elsewhere. Additionally, the HST archives contain WFPC2
images of FCC046 in the filters F814W and F555W. Availability of all
these data was the sole reason for selecting this object.

FCC046 contains a bright nucleus, which is displaced by 1\farcs2 with
respect to the outer isophotes, or $30\%$ of $R_{\rm d}$. The
lopsidedness involves not just the nucleus, but extends to $\sim 2
R_{\rm d}$. Color maps show no traces of large amounts of dust but
suggest the presence of a blue stellar population.
The H$\alpha$ image reveals
ionized emission regions, evidence for ongoing star formation, within
$\sim 1 R_{\rm d}$. The position of the blue stellar population does
not coincide with the lopsidedness of FCC046 and is not likely to be
the cause of it (although it may be a consequence of it). FCC046 is
in the outskirts of the Fornax cluster; its nearest neighbors are
dwarfs over 100 kpc (in projection) away, and the nearest large galaxy
is $> 200$ kpc distant.  Interactions are therefore not likely to
account for its morphology, unless a weakly damped mode has been
excited \citep{w94}.
The kinematics of FCC046 (mean velocity and velocity dispersion) were
extracted from the spectra following the method discussed in \citet{ik03a}.
Despite its E4 flattening, the main body of FCC046 has zero mean rotation.
The lack of rotation excludes the possibility that the eccentric instability
\citep{ars89,ms92,ti98} is responsible for the lopsidedness.

The nucleus is resolved by HST; it has a FWHM$= 0\farcs27$,
significantly larger than the FWHM$= 0\farcs15$ of the stars in the
same image, and therefore cannot be a foreground star.  The nucleus
has the same systemic velocity as the galaxy which rules out a chance
projection of a background object.  It comprises $\sim 10~\%$ of the
$B$-band luminosity of FCC046 and is therefore probably quite massive.
We can use the fact that the outer isophotes of the $N$-body model do
not move appreciably and their center coincides with the model's
center-of-mass (COM) to estimate the relative mass-to-light ($M/L$)
ratios of the disk and nucleus.  In order for the mean position of the
$B$-band light distribution to coincide with the COM, the $M/L$ ratio
of the nucleus has to be less than half that of the disk.  If the disk
consists of an old, metal-poor population with $M/L_B \approx 5
M_\odot/L_{B,\odot}$, this would mean that the nucleus has $M/L_B < 2
M_\odot/L_{B,\odot}$, typical for a very young stellar population (age
$< 2-3$~Gyr). This is consistent with the blue color of the nucleus
and its strong Pa absorption \citep{mddzh03}.  The rotation axis of
the nucleus is not aligned with that of the galaxy's main body,
showing evidence for minor axis rotation with an amplitude of $\approx
5$~km/s.

\begin{figure}
\vspace{6cm}
\special{hscale=44 vscale=44 hsize=570 vsize=170 
         hoffset=-17 voffset=250 angle=-90 psfile="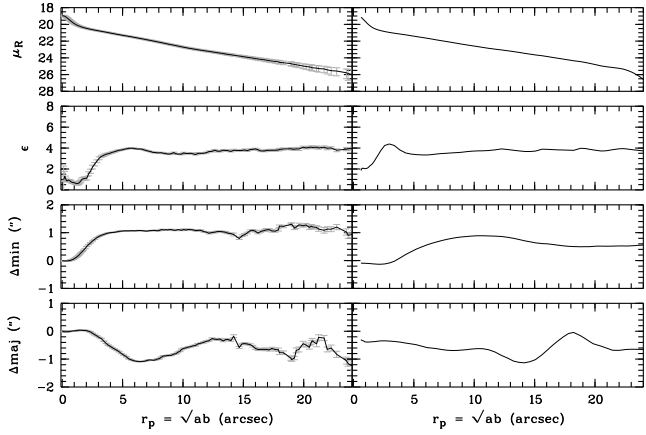"}
\caption{Surface photometry of FCC046 (left panels), compared with the
  $N$-body model (right panels). $a$ and $b$ are respectively the
  semi-major and minor axes of the isophotes. In the top panels, the
  $R$-band surface brightness (in mag/arcsec$^2$) of FCC046 is
  compared with that of the model.  In the next panels, the
  ellipticity $\epsilon = 10(1-b/a)$ of the isophotes is shown. In the
  bottom panels, the position of the center of the isophotes is
  presented. $\Delta {\rm maj}$ and $\Delta {\rm min}$ (in arcseconds)
  quantify the position of the isophotes along the major and minor
  axes, centered on the nucleus.
\label{compsurf}}
\end{figure}

The surface density, the velocity field and the velocity dispersion
field of the $N$-body model at $t=150$ are presented in
Fig. \ref{modelps}. The viewing angle and the spatial scale were
chosen such that the model's surface density gives a fair
approximation to the $B$-band image of FCC046. The nucleus is offset
by $\sim 1''$ to the south-west with respect to the outer isophotes
and, in response, there is an excess in the disk density to the
north-east, as in FCC046. Clearly, the instability engenders a global
response of the whole stellar system. In Fig. \ref{compsurf}, we
compare the surface photometry of FCC046 with the $N$-body model
(convolved with a Gaussian to mimic the seeing conditions of the
observations) in greater detail. Simply shifting the magnitude
zero-point of the $N$-body model and applying the same spatial scale
as in Fig. \ref{modelps} sufficed to match the surface brightness
profile of FCC046. As in FCC046, the isophote centers of the model
oscillate along the major axis; in the model this keeps the total
center-of-mass fixed.  Just outside the nucleus, the density contours
have a distorted shape which, in projection, results in the isophotes
being flatter (around $\sim 5''$ in FCC046 and around $\sim 3''$ in
the model) than at larger radii.  On the whole, the model successfully
reproduces gross features in the surface photometry of FCC046.

\clearpage

\begin{figure}
\vspace{6cm}
\special{hscale=40 vscale=40 hsize=570 vsize=170 
         hoffset=-20 voffset=220 angle=-90 psfile="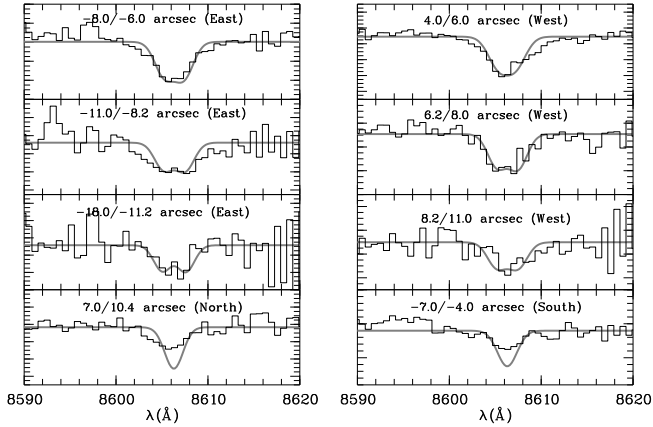"}
\caption{Part of the spectrum of FCC046, around the strongest Ca{\sc
ii} absorption line. The radial bins over which the spectrum has been
summed to boost the $S/N$ are indicated in the figure (left column,
top three panels:~along the major axis, towards east; right column,
top three panels:~along the major axis, towards west; bottom
panels:~along minor axis, towards north (left column) and towards
south (right column)). Radius is measured from the nucleus. Observed
spectrum:~black line; model spectrum:~grey line (see text for
details). This absorption line appears split into two distinct lines,
separated by $70 \pm 25$~km/s, in the outermost major axis
spectra. The $S/N$ in these panels varies from 5-10 to 20-25; adjacent
spectra from the left and right columns have comparable $S/N$. The
minor-axis spectra have the same continuum level as the outermost
major-axis spectra. Clearly, the Ca{\sc ii} line does not appear split
in the minor-axis spectra. The model reproduces rather well the
major-axis line width and the observed line-splitting.
\label{spec}}
\end{figure}

\clearpage

Fig. \ref{modelps} shows that, while the model's mean velocity is
zero, bi-symmetric perturbations of order $\sim 0.06$ are present. The
velocity dispersion profile is asymmetric, rising less rapidly towards
the left side of the model, in the direction of the light excess. The
apparent flattening of FCC046 and its zero mean rotation already hint
at two equal-mass counter-streaming stellar populations. Although at
the limit of what can be done, given the spectral resolution and the
low $S/N$-ratio of the data, the strongest Ca{\sc ii} absorption line
in the major-axis spectrum appears split into two distinct lines (see
Fig. \ref{spec}), separated by $70\pm 25$~km/s; the other, weaker,
lines have much lower $S/N$-ratios. This absorption line is unaffected
by the bright sky OH-emission lines. (If a badly removed sky line were
the cause of the line-splitting, it would also manifest itself in the
center of the galaxy and in the minor axis spectrum, which is not the
case.)  What is particularly compelling about these data is that the
apparent bimodality is present (and absent) where the model predicts
it should be (before the signal disappears in the noise), and that we
see similar splitting in three different spatial bins. To illustrate
this, in Fig. \ref{spec} we plot the observed spectrum and the model
spectrum, consisting of a flat continuum and a very narrow
($\delta$-function) absorption line, broadened with the FORS2
instrumental profile and the model's line-of-sight velocity
distribution (LOSVD), summed over the same spatial bins as the
observed spectrum. We did not use an observed stellar template since
the width of the individual LOSVD peaks is smaller than one FORS2
pixel, making an accurate calculation of a synthetic galaxy spectrum
impossible. Except for underestimating the minor-axis velocity
dispersion, the model reproduces rather well the major-axis line
widths and the observed line-splitting.

\begin{figure}
\vspace{6.25cm}
\special{hscale=40 vscale=40 hsize=570 vsize=180
         hoffset=0 voffset=225 angle=-90 psfile="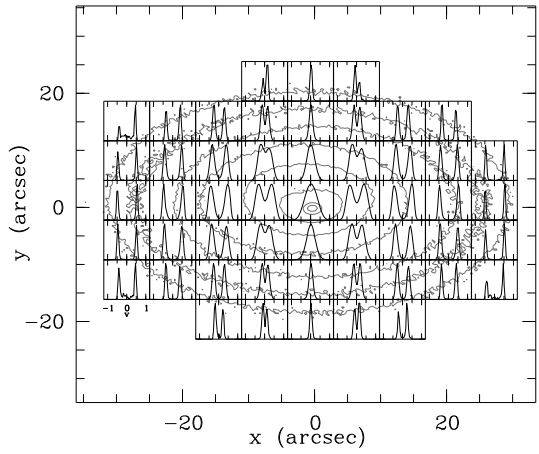"}
\caption{
  A mosaic of the LOSVDs at different points in the projected model (the
  middle panel is centered on the center of the outer isophotes). Big and
  small tickmarks indicate a velocity $v = \pm 1$ and $v = \pm 0.5$,
  respectively, as indicated below the leftmost panel. The isophotes of the
  model, separated by 1~mag/arcsec$^2$, are plotted in grey-scale.
\label{losvd}}
\end{figure}

However, these observations, while very suggestive, need to be
corroborated at higher spectral resolution to resolve the two peaks in
the LOSVDs.  In Fig. \ref{losvd}, we present a mosaic of LOSVDs at
different points in the model. Moving out along the major axis, the
LOSVDs become more double-peaked due to the increasing velocity
separation between the two counter-streaming stellar populations and
the declining velocity dispersion. The small velocity-field
perturbations of Fig.  \ref{modelps} are reflected here in the LOSVDs
being slightly asymmetric.  Although the details of the LOSVD field in
Fig. \ref{losvd} are model-dependent, an observational confirmation of
its salient features would lend considerable support to the hypothesis
that the counter-streaming instability is responsible for off-center
nuclei in dE,Ns.

\clearpage

\clearpage

\section{Discussion}
\label{sec:disc}

If the counter-streaming instability is indeed working in a sizable
fraction of the dE,Ns to drive nuclei off-center, it is natural to ask
why these dEs have counter-rotation in the first place. The first
detections of counter-rotation in galaxy disks \citep{4550,7217} lead
to the hypothesis that the capture of gas with retrograde rotation and
subsequent star-formation produced two counter-rotating stellar disks.
This hypothesis predicts stellar populations of different ages and
metallicities. Generally, the two disks will not have identical
masses, so some net rotation is expected; this is, therefore, an
unlikely explanation for FCC046. 

\citet{tr00} suggested that counter-rotation is produced when a
triaxial halo with an initially retrograde pattern speed slowly
changes to a prograde pattern speed.  However, this situation probably
does not occur very often. Moreover, this mechanism requires evolution
on long timescales $\sim 10^{10}$~yr and hence requires dE,Ns to be
old and unperturbed.

A more likely mechanism is the destruction of box-orbits, which are
the backbone of any slowly rotating triaxial mass-distribution, by a
growing nucleus. Stars on box orbits come very close to the nucleus
and thus can be scattered into loop-orbits; conservation of angular
momentum then requires that direct and retrograde loop orbits be
equally populated \citep{ec94,mq98}.  If the nucleus has a mass larger
than $2\%$ of the total mass of the galaxy, evolution towards an
axisymmetric shape requires a few crossing times ($\approx 10^7 -
10^8$~yr), while our simulations show that the counter-streaming
instability develops on similar timescales. Late infall of gas
\citep{co03} (perhaps as a result of harassment \citep{mo96,ma01}) or
globular clusters \citep{lo01} may be responsible for growing the
nucleus. If this explanation holds, then the two counter-rotating
stellar populations should have equal scale-lengths and chemical
properties and nearly equal mass.  Additionally, the nucleus may
retain some of its initial angular momentum; we note that the nucleus
of FCC046 is indeed rotating.

To summarize:~using $N$-body simulations, we have shown that the
counter-streaming instability is a viable explanation for at least the
dE,N galaxy FCC046. The models can produce offsets as large as
observed and successfully reproduce distinctive features in the
surface photometry of FCC046, such as the oscillation of isophote
centers.  Although at the limit of what can reliably be extracted from
the spectra, we find a tantalizing hint of counter-streaming in the
major-axis spectra of this galaxy, which, however, needs to be
corroborated at higher spectral resolution.

\acknowledgments V.P.D. thanks the Sterrenkundig Observatorium for
hospitality during an early stage of this project. We thank Stelios
Kazantzidis and Anna Pasquali for useful discussions and the referee,
Dr. O. Gnedin, for the suggested improvements.


\begin{thebibliography}{}
  
\bibitem[Adams {\em et al.}(1989)]{ars89} Adams, F.~C., Ruden, S.~P., Shu,
  F.~H. 1989, \apj, 347, 959

\bibitem[Barazza {\em et al.}(2002)]{bbj02} Barazza, F.~D., Binggeli, B., \&
  Jerjen, H. 2002, \aap, 391, 823
   
\bibitem[Binggeli \& Ferguson(1994)]{bf94} Binggeli, B., \& Ferguson, H.~C.
  1994, \aapr, 6, 67
  
\bibitem[Binggeli {\em et al.}(2000)]{bbj00} Binggeli, B., Barazza, F.,
  Jerjen, H. 2000, \aap, 359, 447
  
\bibitem[Conselice {\em et al.}(2003)]{co03} Conselice, C.~J., O'Neil, K.,
  Gallagher, J.~S., Wyse, R.~G. 2003, \apj, 591, 167
  
\bibitem[De Rijcke {\em et al.}(2003a)]{ik03a} De Rijcke, S., Dejonghe, H.,
  Zeilinger, W.~W., \& Hau, G.~K.~T. 2003, \aap, 400, 119
  
\bibitem[De Rijcke {\em et al.}(2003b)]{ik03b} De Rijcke, S., Zeilinger,
  W.~W., Dejonghe, H., \& Hau, G.~K.~T. 2003, \mnras, 339, 225
 
\bibitem[Evans \& Collett(1994)]{ec94} Evans, N.~W., \& Collett, J.~L.
  1994, \apj, 420, L67
 
\bibitem[Geha {\em et al.}(2003)]{ggv03} Geha, M., Guhathakurta, P., \& van
  der Marel, R.~P. 2003, \aj, 126, 1794

\bibitem[Gnedin {\em et al.}(2002)]{g02} Gnedin, O.~Y. , Zhao, H.-S.,
Pringle, J. E., Fall, S.~M., Livio, M., \& Meylan, G. 2002, \apjl,
568, L23

\bibitem[Levison {\em et al.}(1990)]{lds90} Levison, H.~F., Duncan, M.~J., \&
  Smith, B.~F. 1990, \apj, 363, 66
  
\bibitem[Lotz {\em et al.}(2001)]{lo01} Lotz, J.~M., Telford, R., Ferguson,
  H.~C., Miller, B.~W., Stiavelli, M., \& Mack J. 2001, \apj, 552, 572
  
\bibitem[Mayer {\em et al.}(2001)]{ma01} Mayer, L., Governato, F., Colpi, M.,
  Moore, B., Quinn, T., Wadsley, J., Stadel, J., \& Lake, G. 2001, \apj, 559,
  754
  
\bibitem[Merrifield \& Kuijken(1994)]{7217} Merrifield, M.~R., \& Kuijken, K.
  1994, \apj, 432, 575
  
\bibitem[Merritt \& Quinlan(1998)]{mq98} Merritt, D., \& Quinlan, G.~D. 1998,
  \apj, 498, 625

\bibitem[Merritt \& Stiavelli(1990)]{ms90} Merritt, D., \& Stiavelli, M. 1990,
  \apj, 358, 399
  
  
\bibitem[Michielsen {\em et al.}(2003)]{mddzh03} Michielsen, D., De Rijcke,
  S., Dejonghe, H., Zeilinger, W. ~W., \& Hau, G.~K.~T. 2003, \apj, 597, L21
  
\bibitem[Miller \& Smith(1992)]{ms92} Miller, R.~H., \& Smith, B.~F. 1992,
  \apj, 393, 508
  
\bibitem[Moore {\em et al.}(1996)]{mo96} Moore, B., Katz, N., Lake, G.,
  Dressler, A., \& Oemler Jr., A. 1996, Nature, 379, 613
  
\bibitem[Rubin {\em et al.}(1992)]{4550} Rubin, V.~C., Graham, J.~A., \&
  Kenney, J.~D.~P 1992, \apj, 394, L9
  
\bibitem[Palmer \& Papaloizou(1990)]{pp90} Palmer, P.~L., \& Papaloizou, J.
  1990, \mnras, 243, 263

\bibitem[Phillipps {\em et al.}(2001)]{p01} Phillipps, S., Drinkwater,
M., Gregg, M., \& Jones, J. 2001, \apj, 560, 201

\bibitem[Prendergast \& Tomer(1970)]{pt70} Prendergast, K.~H., \&
Tomer, E., 1970, \aj, 75, 67

\bibitem[Rakos \& Schombert(2003)]{rs03} Rakos, K. \& Schombert, J. 2003, accepted for publication by \aj, astro-ph/0312075

\bibitem[Sawamura(1988)]{s88} Sawamura, M. 1988, \pasj, 40, 279
  
\bibitem[Sellwood \& Valluri(1997)]{sv97} Sellwood, J.~A., \& Valluri, M.
  1997, \mnras, 287, 124
  
\bibitem[Sellwood \& Merritt(1994)]{sm94} Sellwood, J.~A., \& Merritt, D. 1994,
  \apj, 425, 530

\bibitem[Sellwood(1983)]{s83} Sellwood, J.~A. 1983, J. Comput. Phys., 50, 337
  
\bibitem[Taga \& Iye(1998)]{ti98} Taga, M., \& Iye, M. 1998, \mnras, 299, 111
  
\bibitem[Tremaine \& Yu(2000)]{tr00} Tremaine, S., \& Yu, Q., 2000, \mnras,
  319, 1

\bibitem[Weinberg(1994)]{w94} Weinberg, M.~D. 1994, \apj, 421, 481
  
\bibitem[Zang \& Hohl(1978)]{zh78} Zang, T.~A., \& Hohl, F. 1978, \apj, 226,
  521

\end{thebibliography}
\end{document}